\documentclass[twocolumn,letter]{jpsj2}

\title{%
Conductance of Disordered Wires with Symplectic Symmetry:
Comparison between Odd- and Even-Channel Cases
}

\author{%
Yositake {\sc Takane}
}

\inst{%
Department of Quantum Matter, Graduate School of Advanced Sciences of Matter,
Hiroshima University, Higashi-Hiroshima 739-8530, Japan
}

\abst{%

The conductance of disordered wires with symplectic symmetry is studied by
numerical simulations on the basis of a tight-binding model on a square
lattice consisting of $M$ lattice sites in the transverse direction.
If the potential range of scatterers is much larger than
the lattice constant, the number $N$ of conducting channels becomes
odd (even) when $M$ is odd (even).
The average dimensionless conductance $\langle g \rangle$ is calculated
as a function of system length $L$.
It is shown that when $N$ is odd, the conductance behaves as
$\langle g \rangle \to 1$ with increasing $L$.
This indicates the absence of Anderson localization.
In the even-channel case, the ordinary localization behavior arises
and $\langle g \rangle$ decays exponentially with increasing $L$.
It is also shown that the decay of $\langle g \rangle$
is much faster in the odd-channel case than in the even-channel case.
These numerical results are in qualitative agreement
with existing analytic theories.
}

\kword{%
symplectic class, tight-binding model, numerical simulation, conductance
}

\begin{document}
\sloppy
\maketitle

The concept of universality classes is important
in considering quantum electron transport
in quasi-one-dimensional disordered wires.~\cite{beenakker}
The universality classes describe transport properties
which are independent of the microscopic details of disordered wires.
Ordinary disordered wires are classified into either of three
universality classes (orthogonal class, unitary class and symplectic class),
called the standard three, according to the symmetries which they possess.
The orthogonal class consists of systems having both time-reversal symmetry
and spin-rotation invariance, while the unitary class is
characterized by the absence of time-reversal symmetry.
The systems having time-reversal symmetry without spin-rotation invariance
belong to the symplectic class.
We here briefly summarize the typical differences
among the standard three classes.
If system length $L$ is much shorter than the localization length $\xi$,
we say that the system is in the metallic regime.
In this regime, the weak-localization (weak-antilocalization) effect
arises in the orthogonal (symplectic) class,
while the effect disappears in the unitary class.
If $L$ is much longer than $\xi$, the conductance decays exponentially
with increasing $L$.
This localization behavior is observed in all the classes.
However, the localization length $\xi$ normalized by the mean free path $l$
is shortest (longest) for the orthogonal (symplectic) class
among the standard three.
That is, the localization behavior is most evident in the orthogonal class.

Exceptional behaviors, which are incompatible with the standard three,
have been found in the electron transport properties of metallic carbon
nanotubes (CNs).~\cite{ando1,ando2,takane1,takane2,takane3,ando3,takane4}
Ando and co-workers~\cite{ando1,ando2} studied CNs of length $L$
when the potential range of scatterers is larger than the lattice constant.
They proved that one perfectly conducting channel is present even when
$L \to \infty$, based on the peculiar symmetry of the reflection matrix.
This indicates the absence of Anderson localization
even in the limit of $L \to \infty$.
It should be noted that a graphite sheet has
symplectic symmetry~\cite{suzuura}
and that the number of conducting channels is always odd
in CNs with long-range scatterers.
Thus, the anomalous behavior should be attributed to
the intrinsic nature of the symplectic class with an odd number of channels.
This indicates that we must separately treat the odd-channel case and
the ordinary even-channel case in considering the symplectic class.
To clarify the nature of the symplectic class with an odd number of channels,
the present author~\cite{takane2} derived the so-called DMPK equation
for the transmission eigenvalues based on a random-matrix theory.
It is shown that the conductance decays exponentially
as $g \to 1$ with increasing $L$.
It is also shown that the decay is characterized by
the typical length scale $\xi_{\rm odd} = (1/4)(N-1) l$,
where $N$ is the number of conducting channels.
Noting that the localization length for the even-channel case
is $\xi_{\rm even} = (2N - 2) l$,~\cite{beenakker}
we observe that the decay of $g$ in the odd-channel case is much faster
than that in the even-channel case.
The same result has been obtained by
a supersymmetric field approach.~\cite{takane4}

The predicted transport properties are so anomalous for
the symplectic class with an odd-number of channels
that further studies based on another independent approach are needed
to reexamine the predictions.
In doing so, an approach which is applicable to both the odd-
and even-channel cases is highly desirable
because it enables us to examine the predicted even-odd differences.
In this letter, we consider a tight-binding model on a square lattice,
which consists of $M$ lattice sites in the transverse direction,
as a model for disordered wires with symplectic symmetry.
We show that if the potential range of scatterers is much larger than
the lattice constant, the number $N$ of conducting channels becomes
odd (even) when $M$ is odd (even).
Thus, we can consider not only the odd-channel case but also
the even-channel cases within this model.
We calculate the ensemble-averaged dimensionless conductance
$\langle g \rangle$ as a function of system length $L$.
It is shown that when $N$ is odd, the conductance behaves as
$\langle g \rangle \to 1$ with increasing $L$.
In the even-channel case, the ordinary localization behavior arises
and $\langle g \rangle$ vanishes with increasing $L$.
It is also shown that the decay of $\langle g \rangle$ with increasing $L$
is much faster in the odd-channel case than in the even-channel case.
These numerical results are qualitatively in agreement with the DMPK equation.

As a microscopic model for disordered wires with symplectic symmetry,
we consider the following tight-binding model on a square lattice,
which is infinitely long in the longitudinal direction
and consists of $M$ lattice sites in the transverse direction,
\begin{align}
    \label{eq:hamiltonian}
  H & = \sum_{j= -\infty}^{\infty} \sum_{l=1}^{M}
        \sum_{\sigma, \sigma' = \uparrow, \downarrow}
              \nonumber \\
    & \hspace{10mm} \times
        \bigl\{ \lvert j,l+1, \sigma \rangle
        \hat{V}_{x}(\sigma, \sigma')
        \langle j,l, \sigma' \rvert
                        + {\rm h. c.} \bigr\}
             \nonumber \\
    & \hspace{3mm}
      + \sum_{j= -\infty}^{\infty} \sum_{l=1}^{M}
        \sum_{\sigma, \sigma' = \uparrow, \downarrow}
              \nonumber \\
    & \hspace{10mm} \times
        \bigl\{ \lvert j+1,l, \sigma \rangle
        \hat{V}_{y}(\sigma, \sigma')
        \langle j,l, \sigma' \rvert
                       + {\rm h. c.} \bigr\}
            \nonumber \\
    & \hspace{3mm}
      + \sum_{j= -\infty}^{\infty} \sum_{l=1}^{M}
        \sum_{\sigma = \uparrow, \downarrow}
        \lvert j,l, \sigma \rangle
        V_{\rm imp}(j,l) \langle j,l, \sigma \rvert
\end{align}
with
\begin{align}
   \hat{V}_{x} & = \left( \begin{array}{cc}
                       0 & - {\rm i}t_{x} \\
                       - {\rm i}t_{x} & 0 \\
                \end{array} \right) ,
              \\
   \hat{V}_{y} & = \left( \begin{array}{cc}
                       0 &   t_{y} \\
                       - t_{y} & 0 \\
                \end{array} \right) .
\end{align}
Here, $V_{\rm imp}(j,l)$ denotes the impurity potential at the $(j,l)$th site.
Note that our Hamiltonian has full symplectic symmetry and
can be viewed as a discrete version of the effective-mass Hamiltonian
for a graphite sheet if $t_{x} = t_{y}$.~\cite{ando1}
It is equivalent to Ando model~\cite{ando4}
in the strong spin-orbit interaction limit.
We impose the periodic boundary condition in the transverse direction.
That is, $\lvert j,M+1, \sigma \rangle = \lvert j,1, \sigma \rangle$.
We hereafter assume that the lattice constant is equal to unity.

As a function of the longitudinal wave number $q$ ($- \pi < q \le \pi$),
the dispersion relation is
\begin{equation}
  E_{n}(q) = \pm 2 t_{y}
             \sqrt{ \sin^{2}q
                    + r^{2} \sin^{2} \left( \frac{2\pi n}{M} \right) } ,
\end{equation}
where $r = t_{x}/t_{y}$,
$n = 0, \pm 1, \cdots , \pm (M-1)/2$ for an odd $M$
and $n = 0, \pm 1, \cdots , \pm (M-2)/2, M/2$ for an even $M$.
We hereafter consider only the subbands with $E_{n}(q) > 0$.
We see that there are two valleys:
one located at $q = 0$ and the other at $q = \pm \pi$.
Let us call the former one valley A and the latter one valley B.
We consider the odd-$M$ case.
The nondegenerate lowest subband with $n = 0$ provides
a conducting channel for each valley
if the Fermi energy $E_{\rm F}$ satisfies $0 < E_{\rm F} < 2t_{y}$.
The other subbands are two-fold degenerate.
This means that each subband provides two conducting channels
for each valley if it crosses the Fermi level.
Thus, the total number $N_{\rm total}$ of conducting channels
is equal to $2, 6, 10, 14, 18, \cdots$ depending on $E_{\rm F}$.
In the even-$M$ case, the lowest subband corresponds to $n = 0$ and $M/2$
and is thus two-fold degenerate.
Furthermore, the other subbands are four-fold degenerate in this case.
Consequently, $N_{\rm total} = 4, 12, 20, \cdots$.

The above argument tells us that
the total number of conducting channels is always even.
However, there is a simple way to effectively realize the odd-channel case.
The following way is inspired by the observation
that the odd-channel case can be realized in CNs if the potential range of
scatterers is larger than the lattice constant.~\cite{ando1,ando2}
We focus our attention on the case where $E_{\rm F}$ is very small,
that is, $0 < E_{\rm F} \ll 2t_{y}$.
In this case,
$q$ for any conducting channels in valley A becomes very small,
while that in valley B is nearly equal to $\pm \pi$.
This means that intervalley scattering must be accompanied by
a large momentum transfer.
Such a scattering is suppressed when the spatial range of scatterers
is much larger than the lattice constant.
Thus, valley A and valley B effectively decouple each other
if only such scatterers are present.
This means that if $M$ is odd,
we can realize a disordered wire with symplectic symmetry
having an odd number of conducting channels.
Let $N$ be the number of conducting channels for each valley.
$N = 1, 3, 5, 7, 9, \cdots$ depending on $E_{\rm F}$
in the odd-$M$ case, while $N = 2, 6, 10, \cdots$ in the even-$M$ case.
Thus, we can examine the even-odd differences within a single model.

We assume that the impurity potential at the site $(j,l)$ arising from
a scatterer at the site $(j_{0},l_{0})$ is given by
\begin{equation}
  V_{(j_{0},l_{0})}(j,l)
     = w_{(j_{0},l_{0})}
       \exp \left( - \frac{(j-j_{0})^{2}}{\Lambda_{y}^{2}}
                   - \frac{(l-l_{0})^{2}}{\Lambda_{x}^{2}}
            \right) .
\end{equation}
To suppress the intervalley scattering,
$\Lambda_{y}$ must be larger than the lattice constant.
We randomly distribute scatterers in the finite region of $1 \le j \le L$.
Thus, $V_{\rm imp}(j,l)$ in the Hamiltonian is given by
\begin{equation}
   V_{\rm imp}(j,l)
    = \sum_{j_{0}= 1}^{L} \sum_{l_{0}= 1}^{M}
      V_{(j_{0},l_{0})}(j,l) .
\end{equation}
The amplitude $w_{(j,l)}$ is distributed uniformly within the range of
$-W/2 < w_{(j,l)} < W/2$.
Let $p$ be the probability that each site is occupied by a scatterer.
The strength of the impurity potential is controlled by $W$ and $p$.
We consider that the length of our system is equal to $L$.

The number of conducting channels in each valley is $N$,
so that the total dimensionless conductance is given by
\begin{equation}
      \label{eq:landauer}
   g_{\rm total}
      = \sum_{a, b = 1}^{N} \sum_{{\rm X}, {\rm Y} = {\rm A}, {\rm B}}
                \lvert t_{{\rm X}a,{\rm Y}b} \rvert^{2} ,
\end{equation}
where $t_{{\rm X}a,{\rm Y}b}$ is the transmission coefficient
for an electron coming from the $b$th channel in valley Y
and going to the $a$th channel in valley X.
If the two valleys are completely decoupled,
eq.~(\ref{eq:landauer}) is reduced to
\begin{equation}
   g_{\rm total}
      = 2 \sum_{a, b = 1}^{N} \lvert t_{{\rm A}a, {\rm A}b} \rvert^{2} .
\end{equation}
Expecting that a good decoupling is achieved,
we focus on the dimensionless conductance $g$ for each valley,
which is defined as $g \equiv g_{\rm total}/2$.
We can numerically obtain $t_{{\rm X}a,{\rm Y}b}$
by a recursive Green-function technique.
We calculate the ensemble average $\langle g \rangle$
for the case of $M = 17$ as an example of the odd-channel case
and that for $M = 16$ as an example of the even-channel case.
In the following numerical calculations, we set $r = 0.1$, $\Lambda_{y} = 4$,
$\Lambda_{x} = 1$, $W/t_{y} = 0.7$ and $p = 0.2$.
The ensemble average is taken over 1000 samples for all the cases.
We here present the reason why we have set $r = 0.1$.
As noted above, $E_{\rm F}$ should satisfy $0 < E_{\rm F} \ll 2t_{y}$
to suppress the intervalley scattering.
If $M$ is very large, we can realize a given channel number $N$
for a sufficiently small $E_{\rm F}$.
However, if $M$ is not large, the condition $0 < E_{\rm F} \ll 2t_{y}$
cannot be satisfied for a given $N$
and thereby the intervalley scattering is relatively enhanced.
In such a situation, we can suppress it by reducing $r$.
Note that with decreasing $r$, the lowest subband does not change,
while the dispersion for each higher subband goes down towards
that for the lowest subband.
Thus, a given $N$ can be achieved by a relatively small $E_{\rm F}$
compared with the isotropic case of $r = 1$.

\begin{figure}[htb]
\begin{center}
\includegraphics[height=8cm]{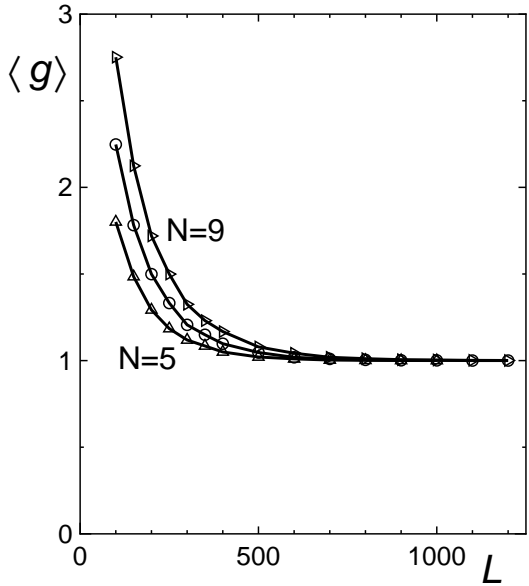}
\end{center}
\caption{The dimensionless conductance $\langle g \rangle$
for $N = 5, 7$ and $9$ as a function of system length $L$.
}
\end{figure}

\begin{figure}[htb]
\begin{center}
\includegraphics[height=8cm]{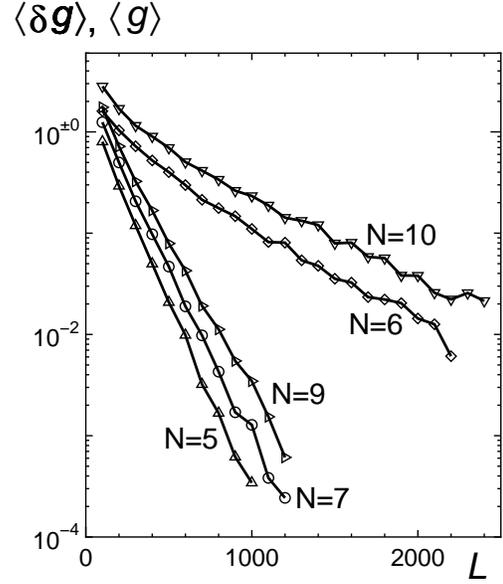}
\end{center}
\caption{The deviation $\langle \delta g \rangle$ for $N = 5, 7$ and $9$
and the dimensionless conductance $\langle g \rangle$ for $N = 6$ and $10$
as a function of system length $L$.
}
\end{figure}
The ensemble-averaged dimensionless conductance $\langle g \rangle$
for the odd-channel case is displayed in Fig.~1 as a function of $L$.
The Fermi energies are chosen as $E_{\rm F}/t_{y} = 0.08, 0.11$ and $0.14$,
which correspond to $N = 5, 7$ and $9$, respectively.
We observe that $\langle g \rangle$ decreases towards unity
with increasing $L$.
This result is in agreement with the DMPK equation
which predicts the conductance quantization
(i.e., $g = 1$) in the long-$L$ limit.
The deviation $\langle \delta g \rangle \equiv \langle g \rangle - 1$ from
the quantized value is displayed in Fig. 2.
We observe that $\langle \delta g \rangle$ decays exponentially
with increasing $L$
and that the decay becomes faster with decreasing $N$.
For the even-channel case, we calculated the averaged dimensionless
conductance $\langle g \rangle$ for $E_{\rm F}/t_{y} = 0.08$ and $0.15$,
which correspond to $N = 6$ and $10$, respectively,
and display the results in Fig.~2.
The behavior of the conductance in the even-channel case
is very similar to that of $\langle \delta g \rangle$.
It decays exponentially with increasing $L$
and the decay becomes faster with decreasing $N$.
This tendency is in agreement with the DMPK equation.~\cite{takane2}

Now we compare
the behavior of $\langle \delta g \rangle$ in the odd-channel case
and that of $\langle g \rangle$ in the even-channel case.
From Fig. 2, we observe that the decay of $\langle g \rangle$
is much slower than that of $\langle \delta g \rangle$.
This tendency is also in agreement with the DMPK equation.
Let us estimate characteristic length scales.
The $L$-dependence of $\langle \delta g \rangle$ is assumed to be
\begin{equation}
  \langle \delta g \rangle \sim {\rm e}^{- \frac{L}{2\xi_{\rm odd}}} ,
\end{equation}
where we call $\xi_{\rm odd}$ the conductance decay length.
For the even-channel case, we assume that
\begin{equation}
  \langle g \rangle \sim {\rm e}^{- \frac{L}{2\xi_{\rm even}}} ,
\end{equation}
where $\xi_{\rm even}$ is the localization length.
The DMPK equation predicts~\cite{takane2}
\begin{equation}
  \xi_{\rm odd}(N)  = \frac{1}{4}(N-1) l ,
\end{equation}
while~\cite{beenakker,mello1,marcedo}
\begin{equation}
  \xi_{\rm even}(N) = (2N - 2) l ,
\end{equation}
where $l$ is the mean free path.
From the above equations, we find that
\begin{equation}
  \frac{\xi_{\rm even}(10)}{\xi_{\rm odd}(9)} = 9 .
\end{equation}
We obtain from our numerical results that
$\xi_{\rm even}(10) \approx 260$ and $\xi_{\rm odd}(9) \approx 70$,
which result in $\xi_{\rm even}(10)/\xi_{\rm odd}(9) \approx 3.7$.
Thus, our numerical results are not in quantitative agreement with
the DMPK equation, although we have obtained the qualitative agreement.
The reason for the quantitative discrepancy is not clear at this stage,
however, it may be related to vertex corrections.
Suzuura and Ando~\cite{suzuura} have calculated the weak-antilocalization
correction in a graphite sheet by a diagrammatic perturbation theory and
have shown that vertex corrections are important for quantitative argument.
In contrast, the derivation of the DMPK equation is based on
the assumption of equivalent scattering channels,~\cite{takane2,mello2}
within which such corrections vanish.
It is very interesting to study the influence of vertex corrections
on $\langle g \rangle$ in the long-$L$ regime,
however, this is difficult.

In summary, we have considered a tight-binding model on a square lattice
as a model for disordered wires
which belong to the symplectic universality class.
We have shown that if the potential range of scatterers is much larger than
the lattice constant, this model enables us to study
both the odd- and even-channel cases.
We have calculated the ensemble-averaged dimensionless conductance
$\langle g \rangle$ as a function of system length $L$.
It is shown that the conductance in the odd-channel case behaves as
$\langle g \rangle \to 1$ with increasing $L$,
while $\langle g \rangle \to 0$ in the even-channel case.
It is also shown that the decay of $\langle g \rangle$
with increasing $L$ is much faster in the odd-channel case
than in the even-channel case.
These numerical results are qualitatively in agreement with the DMPK equation.
However, for the typical length scales
which characterize the decay of $\langle g \rangle$,
we have not obtained a quantitative agreement
between the numerical results and the DMPK equation.

\end{document}